\documentclass[journal]{IEEEtran}
\usepackage{amsmath,amsfonts}
\usepackage{algorithmic}
\usepackage{array}
\usepackage{graphicx}
\usepackage[caption=false,font=normalsize,labelfont=sf,textfont=sf]{subfig}
\usepackage{url}
\usepackage{verbatim}
\usepackage{color}
\usepackage{booktabs}
\usepackage[para]{threeparttable}
\usepackage{xcolor}
\usepackage{ulem}
\usepackage{comment}
\usepackage[colorlinks=true, allcolors=blue]{hyperref}
\usepackage{hyperref}
\hypersetup{
    colorlinks=true,
    linkcolor=blue,
    filecolor=magenta,      
    urlcolor=cyan,
    pdftitle={Sharelatex Example},
    bookmarks=true,
    pdfpagemode=FullScreen,
    }

\def\BibTeX{{\rm B\kern-.05em{\sc i\kern-.025em b}\kern-.08em
    T\kern-.1667em\lower.7ex\hbox{E}\kern-.125emX}}
\usepackage{balance}

\begin{document}
\bstctlcite{IEEEexample:BSTcontrol}

\title{Skipper-in-CMOS: Non-Destructive Readout with Sub-Electron Noise Performance for Pixel Detectors}

\author{Agustin J. Lapi, 
Miguel Sofo-Haro, 
Benjamin C. Parpillon, 
Adi Birman, 
Guillermo Fernandez-Moroni, 
Lorenzo Rota, 
Fabricio Alcalde Bessia, 
Aseem Gupta, 
Claudio R. Chavez Blanco, 
Fernando Chierchie, 
Julie Segal,
Christopher J. Kenney, 
Angelo Dragone, 
Shaorui Li, 
Davide Braga, 
Amos Fenigstein,
Juan Estrada, 
Farah Fahim
\thanks{This work was funded by the DOE Office of Science Research Program for Microelectronics Codesign through the project “Hybrid Cryogenic Detector Architectures for Sensing and Edge Computing enabled by new Fabrication Processes” (LAB 21-2491), with support from DOE's KA25 Advanced Technology R\&D program.

}

\thanks{A. Lapi is with the Instituto de Inv. en Ing. El\'ectrica ``Alfredo Desages'' (IIIE), CONICET, with the Universidad Nacional del Sur (UNS), Bah\'ia Blanca 8000, Argentina, and with the Fermi National Accelerator Laboratory, Batavia, IL, USA (e-mail: lapiagustinjavier@gmail.com).}

\thanks{C. Chavez Blanco is with Facultad de Ingenier\'ia de la Universidad Nacional de Asunci\'on (UNA), San Lorenzo, Paraguay, with Universidad Nacional del Sur (UNS), Bah\'ia Blanca, Argentina, and with Fermi National Accelerator Laboratory, Batavia, IL, USA.}
\thanks{F. Chierchie is with Instituto de Inv. en Ing. El\'ectrica ``Alfredo Desages'' (IIIE), CONICET and Universidad Nacional del Sur (UNS), Bah\'ia Blanca, Argentina.}
\thanks{D. Braga, J. Estrada, F. Fahim, S. Li, B. C. Parpillon are with Fermi National Accelerator Laboratory, IL, USA.} 
\thanks{G. Fernandez-Moroni is with Fermi National Accelerator Laboratory and Department of Astronomy and Astrophysics, University of Chicago, IL, USA.}
\thanks{A. Gupta, C. J. Kenney, J. Segal, L. Rota, and A. Dragone are with SLAC National Accelerator Laboratory, CA, USA.}
\thanks{F. Alcalde Bessia is with Instituto de Nanociencia y Nanotecnolog\'ia (CNEA-CONICET), San Carlos de Bariloche, Argentina.}
\thanks{M. Sofo-Haro is with Universidad Nacional de C\'ordoba (UNC), CONICET and CNEA, C\'ordoba, Argentina.}
\thanks{A. Birman and A. Fenigstein are with Tower Semiconductor, Migdal Haaemek, Israel.}}

\maketitle

\begin{abstract}
The Skipper-in-CMOS image sensor integrates the non-destructive readout capability of Skipper Charge Coupled Devices (Skipper-CCDs) with the high conversion gain of a pinned photodiode in a CMOS imaging process, while taking advantage of in-pixel signal processing.
This allows both single photon counting as well as high frame rate readout through highly parallel processing.
The first results obtained from a $15 \times 15$ $\mu$m\textsuperscript{2} pixel cell of a Skipper-in-CMOS sensor fabricated in Tower Semiconductor's commercial 180 nm CMOS Image Sensor process are presented. Measurements confirm the expected reduction of the readout noise with the number of samples down to deep sub-electron noise of 0.15$\rm e^-$, demonstrating the charge transfer operation from the pinned photodiode and the single photon counting operation when the sensor is exposed to light. 
The article also discusses new testing strategies employed for its operation and characterization.
\end{abstract}


\section{Introduction}
Skipper-CCDs have an output readout stage with a floating gate that allows multiple non-destructive sampling of the charge packet for each pixel. After averaging the multiple samples, it is possible to achieve an extremely low readout noise of 0.068\,$e_{rms}^-$/pixel, reaching the absolute theoretical limit of silicon of 1.1\,eV in energy threshold. This technology has been recently proved in \cite{Tiffenberg:2017aac}, and its development has been motivated to provide the technology needed to build the next generation of dark matter and neutrino experiments that will be at the forefront of exploring physics beyond the Standard Model \cite{Barak2020}\cite{oscura2022}\cite{CONNIE2019}\cite{violeta2020}. The \textit{Sub-Electron Noise Skipper-CCD Experimental Instrument} (SENSEI) has produced world-leading constraints on low mass dark matter searches \cite{Abramoff:2019dfb}. The extremely low readout noise of Skipper-CCD allows the detection of single photons in the optical and near-infrared range. Unlike other silicon detectors with an avalanche gain, with Skipper-CCDs it is possible to directly count the exact number of electrons per pixel and therefore the number of photons that interacted on each pixel, being only limited by the Fano noise \cite{janesick_1988}\cite{Rodrigues_2020}. Skipper-CCDs have therefore been identified as a powerful tool for quantum information science, giving access to entangled measurements in momentum and spatial variables for single photons and motivating their use for new physics searches \cite{estrada_dark_photon}.

Skipper-CCDs are fabricated in a dedicated facility using a customized process for scientific CCDs \cite{ccdProcess}. This process is required to produce the overlapping of the gates structures needed to achieve high charge transfer efficiency between pixels. Due to the very low demand of scientific CCDs, compared to commercial CMOS imagers, the number of facilities dedicated to scientific CCDs has reduced to only a few in the world today, and this number is expected to continue dropping \cite{dawson2019maintaining}. 
On the other hand, imagers fabricated in CMOS process have dominated the market of high-demand consumer cameras, and therefore several fabrication facilities with many processing options are available.

Advances in the CMOS technology fabrication processes have enabled the implementation of pixels with higher sensitivity. For example, the Jots have a very high charge-to-voltage conversion gain and a vertical PPD, resulting in the first demonstration of deep sub-electron read noise (DSERN) in CMOS sensors at room temperature \cite{Quanta-jot, quanta}.
These small structures enabled single photon detection, reaching 1000fps and many of them are grouped to form pixels with larger photon collection capacity \cite{review_of_quanta}. Other authors have reported low noise (0.22-2$\rm e^-$) with rates in the order of 50-100fps \cite{ma20154mp}\cite{fowler20105}.
Another type of single quanta image sensor is based on the Single Photon Avalanche Diode (SPAD) \cite{morimoto_spad_ll, Ogi_spad}, which relies on the multiplication gain of the collected photoelectron to create a high signal larger than the single electron equivalent noise. They usually have a large pixel pitch as a need to separate the pixels due to the use of high electric fields to trigger avalanches. A different approach is presented in this paper, which is based on a PD connected to a floating gate readout structure, which allows the nondestructive readout used in Skipper-CCD to be implemented in each pixel. This architecture allows the arbitrary reduction of the readout noise at the expense of extra readout time.

There are therefore a number of compelling reasons to implement Skipper-CCDs in CMOS. The idea can be traced to \cite{fossum1993active}\cite{mendis1993design}, and was more recently revisited in \cite{stefanov2020simulations}\cite{zhou2022low}. 
While previous works have successfully implemented CCDs in different single-poly CMOS fabrication technologies achieving high charge transfer efficiencies \cite{boulenc2017high}\cite{crooks2013kirana}\cite{marcelot2014study}\cite{fife2009design}, this article demonstrates the operation of non-destructive charge readout of a n-channel Skipper-in-CMOS pixel for image sensors fabricated in Tower Semiconductor TS18 180nm fabrication process.
This effort aims to demonstrate low-noise imagers for single-photon detection, leveraging state-of-the-art commercial CMOS processes for ease of manufacture, but also to allow massively parallel readout by hybrid integration with a dedicated readout ASIC including analog-to-digital parallel processing and high bandwidth data transmission. The development could provide the next generation of detectors for future dark-matter and neutrino experiments, astronomy, and quantum imaging that require single-photon resolution and high readout speed. 

The next section describes the architecture of the characterized pixel. Section \ref{sec: experimental results} addresses first the noise performance of the readout stage and the charge transfer efficiency. Then, the structure is exposed to light to evaluate the charge transfer from the active volume of the pixel and to show the single photon counting capability. A discussion on future work and outlook concludes the article.

\section{Pixel architecture}
\begin{figure}[h]
    \centering
    \includegraphics[page=1,width=\linewidth]{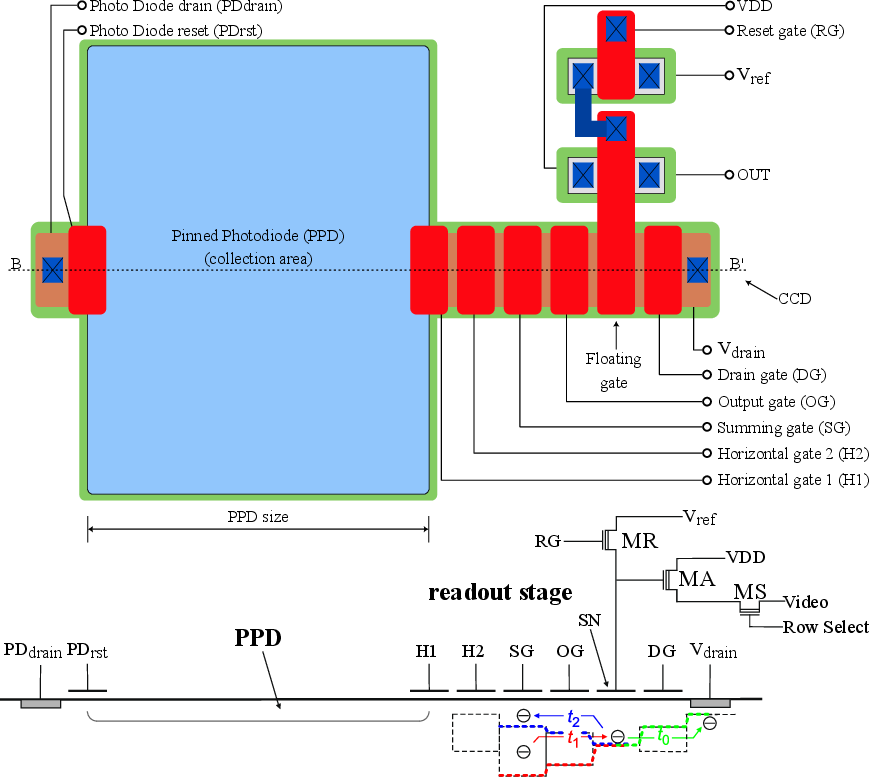}
    \caption{At the top, the top view of the pixel. Dashed line indicates bottom schematic view. The dashed lines in the bottom plot show qualitative voltage potential in the channel of the output stage. The different colors give the status of the potential of the involved gates during charge movement for the different time instants $t_0$, $t_1$, and $t_2$. The spacing of the CCD gates is 250nm.}
    \label{fig:CMOS-gate-structure}
 \end{figure}

A top-view and cross-section drawing of the pixel with its critical components marked is presented in Fig.~\ref{fig:CMOS-gate-structure}. The primary active region for the collection of the ionized electrons is the Pinned Photo Diode (PPD), on the left it is connected via the PD reset (PD${\text{rst}}$) gate to the photodiode drain (PD${\text{drain}}$). The main purpose of this path is to discharge the PPD and reset it to a known reference voltage. 
 ~%
 \begin{figure}
    \centering
    \includegraphics[page=1,width=\linewidth, trim= {0 0 0 0}, clip]{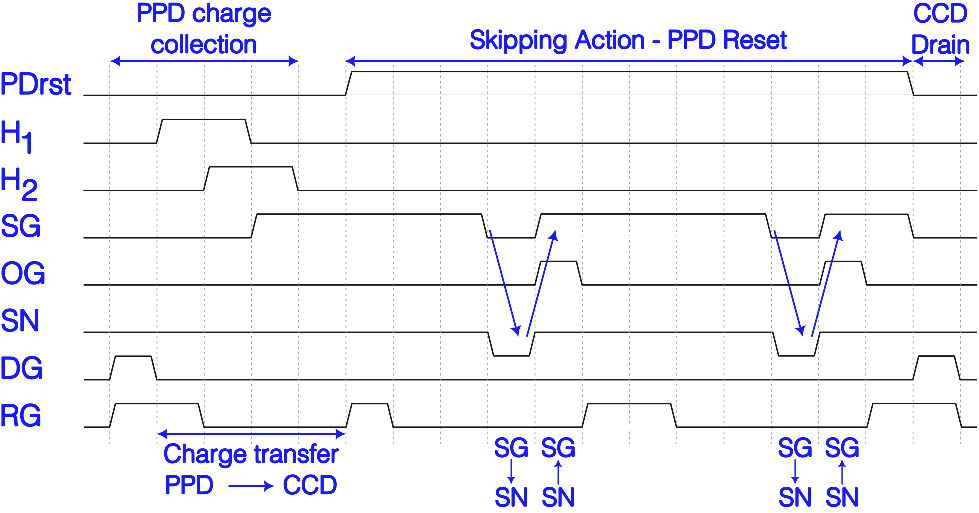}
    \caption{Timing diagram and charge transfer dynamic for the Skipper-in-CMOS pixel. The CCD gates (SG, OG, SN, DG) are operated as a 3-phase CCD.}
    \label{fig:TimingDiagram}
 \end{figure}~%
 
On the right side, the PPD is capacitively connected to two transfer gates, denoted as H1 and H2. H1 and H2 work in tandem either to isolate the PPD from the output stage (when set in low voltage) or to create a potential gradient, helping the charge move from the PPD toward the output stage, specifically to the Summing Gate (SG). The series SG and Output Gate (OG) are responsible for the back-and-forth charge transfer to the floating Sense Node (SN). Finally the Dump Gate (DG) and the V$_{\text{drain}}$ are located to the right of the SN. The timing diagram for operating the pixel is shown in Fig~\ref{fig:TimingDiagram}, involving the following phases.

\textit{Initial reset phase:} Resetting the pixel removes the previous charge packet from the PPD. Subsequently, SN is reset to V$_{\text{ref}}$ through the MR transistor using the Reset Gate (RG) (at time $t_0$ in Fig.~\ref{fig:CMOS-gate-structure}). The readout system then measures this reference potential known as the pedestal level. 

\textit{Charge transfer to the summing node SG}, via the transfer gates H1 and H2.

\textit{Charge sensing and skipping:} The back-and-forth charge transfer occurs between the summing and sensing nodes SG and SN (at $t_1$ and $t_2$ in Fig.~\ref{fig:CMOS-gate-structure}). 
The charge packet moves to SN by lowering SG and OG below V$_{\text{ref}}$, establishing a barrier between the SG and SN, similar to the standard Skipper-CCD operation \cite{skipper_2012}.
The new potential value of the SN is the signal level. A Dual Slope Integration (DSI) technique \cite{janesick2001scientific} is used by the readout system to measure the charge in the pixel by subtracting the signal level from the pedestal level.
Following the completion of the charge measurement, non-destructive readout can be performed by transitioning OG and SG to high voltage levels, allowing the charge to be drawn back underneath the SG. 

The voltage signal in the SN is read out via the n-type source follower transistor MA, designed and optimized for low noise performance, with the assistance of the row-select transistor MS. In the interim periods between charge measurements, SN is reset to the V$_{\textbf{ref}}$ potential through the MR reset transistor.

\textit{CCD reset:} Eventually, after several charge measurements, when no further sampling is required, DG is raised high, prompting the movement of the charge to the ohmic contact at V$_{\text{drain}}$, connected to an external bias voltage.
\begin{figure}[h!]
    \centering
    \includegraphics[page=1,width=\linewidth]{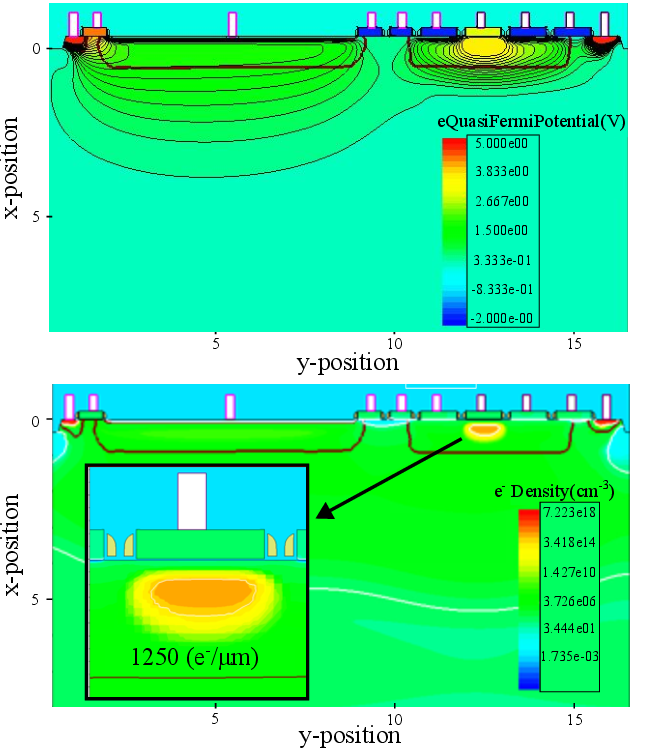}
    \caption{2D simulation of the pixel after the transfer of the charge packet (under the output gate OG), for the potential at the top, and the electron density at the bottom. In the latter, the zoomed-in view of the output gate shows the concentrated charge packet.}
    \label{fig:cmos-simulation}
 \end{figure}
Designing the pixel at the device level necessitates the optimization of each component individually through layout and implant schemes tailored to its specific role. The readout source follower MA should be optimized for low noise using a combination of device implant and size, favoring larger sizes for improved performance. However, this comes with a tradeoff, as the capacitance of MA affects the conversion gain of SN. Despite being optimized for noise, a surface channel is employed to achieve higher gain.

The length of DG requires optimization to ensure the isolation of the read node from the drain. H1 and H2 should exhibit perfect transfer, which, in turn, trades off with the isolation of the PPD during integration to prevent the leakage of collected electrons into the CCD device. Optimization of CCD phases, including SG, OG, and SN, is crucial for various performance parameters, with trade-offs between full well (FW) and conversion gain. A larger FW is associated with higher capacitance and lower conversion gain.

To prevent parasitic light sensitivity of the CCD during multiple read stages, a p-type implant under the CCD is essential. However, this poses a challenge to charge transfer efficiency (CTE), even in a short CCD register with multiple read activities. The use of a buried channel CCD becomes crucial for maintaining CTE, ensuring no loss of charge occurs during the repeated movement of charge between the CCD phases. Detailed TCAD simulations were conducted to optimize these requirements.

Fig.~\ref{fig:cmos-simulation} presents 2D simulations of the pixel, illustrating equipotential curves (top) and electron density (bottom) for charge already transferred to the CCD phase. The buried channel filled with charge is distinctly observed in the simulation results.

\section{Experimental results}
\label{sec: experimental results}
\begin{figure}[h]
    \centering
    \includegraphics[page=1,width=0.8\linewidth, trim={0 0 0 0cm}, clip]{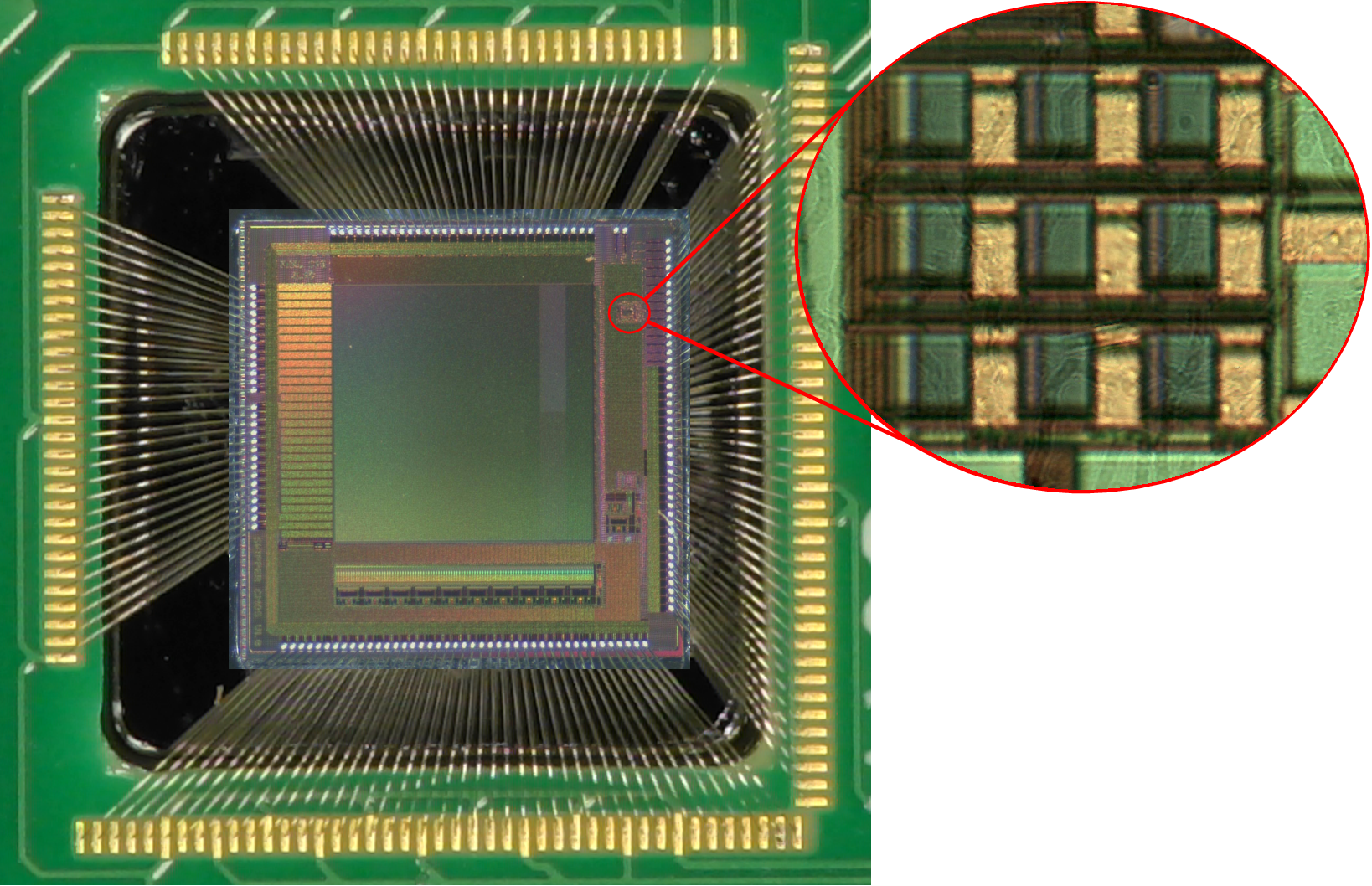}
    \caption{$5 \times 5$ mm Skipper-in-CMOS prototype packaged at Fermilab. While the performance of the $200 \times 200$ pixel matrix is currently under study, this article exclusively showcases results obtained from the central pixel of the test structures highlighted in the zoomed-in view, which contains a $3 \times 3$ matrix of 15 $\mu$m pixels.}
    \label{fig:cmos-packaged}
 \end{figure}

A $5 \times 5$ mm\textsuperscript{2} front side illuminated (FSI) Skipper-in-CMOS prototype chip, bonded to a PCB board is shown in Fig.~\ref{fig:cmos-packaged}. The chip includes a matrix of $200 \times 200$ pixels of 15 $\mu$m pitch, along with additional test structures. The inset image shows a smaller test structure matrix of $3 \times 3$ pixels. While the test measurement of the large main pixel matrix is currently underway, the central pixel within the smaller matrix was extensively characterized and its detailed results are presented in this paper.

\begin{figure}[h!]
    \centering
    \includegraphics[page=1,width=\linewidth]{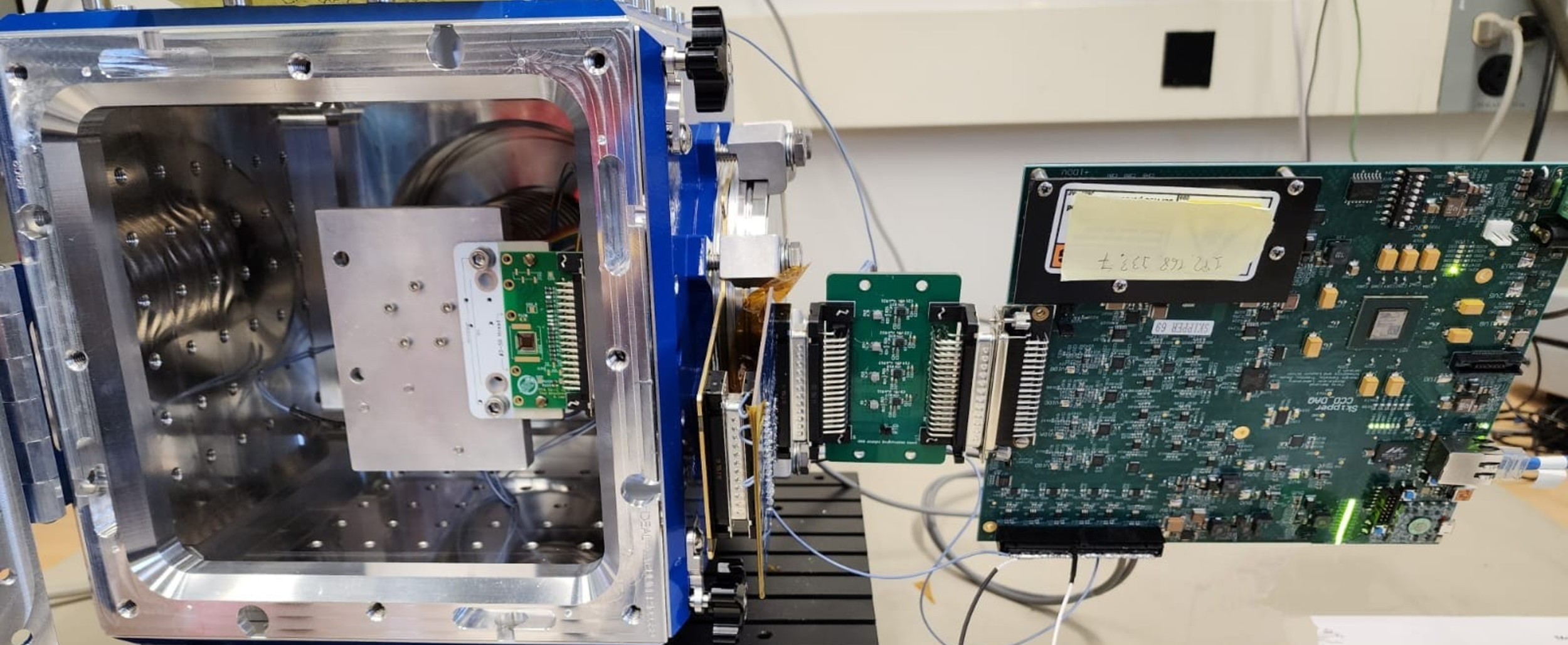}
    \caption{Experimental setup, where the Skipper-in-CMOS can be seen attached to a cold plate, connected through a cable to an interface board and the LTA readout controller outside of the cryocooler.}
    \label{fig:setup}
 \end{figure}

The detector is glued to a silicon substrate, which is then securely attached to the PCB. The silicon substrate possesses identical thermal conductivity, mitigating the risk of die breakage during thermal cycles. The chip is positioned within a square cutout and wire bonded to the PCB. The experimental configuration, as depicted in Fig.~\ref{fig:setup}, involves housing the sensor in a vacuum-sealed, light-tight dewar, positioned on a cold plane connected to a cryocooler and a temperature control system. A constant temperature of 118 K was selected to minimize dark current generation and reduce single electron generation, which facilitates the pixel measurement in low light conditions and enables operation in the single photon regime. Within the controlled chamber, a strategically placed LED illuminates the sensor from the front side, ensuring controlled light exposure.

The PCB sensor board is connected through a flex cable to a feedthrough connector as seen in Fig.~\ref{fig:setup}. Discrete analog electronics bias the MA amplifier in a source follower configuration with $15\mu$A of biasing current. The resulting output video signal is fed to a front-end electronic board, which is connected to the Low Threshold Acquisition (LTA) readout system (both of these are referenced in \cite{cancelo2021low}). The LTA provides clocks and biases for the sensor, digitizes the video signal, and performs digital post-processing to obtain the pixel value through a digital Double Sampling Integrator (DSI).

To bias the MR transistor, a readout system clock signal is used. Applying a small square signal (V$_{\text{ref}}$) around the polarization point enables verification of the correct biasing of the output transistor MA, achieved by examining the output raw video signal. This process allows for the measurement of the gate gain from the Sense Node (SN) to the input of the ADC in the readout controller (located at the end of the video chain). The sensitivity of the node in ADUs/$e^-$ is instead calibrated in absolute terms based on the single electron peaks. Additionally, the LTA controls the LED in sync with the readout sequence. The flexibility offered by this process makes it an important feature of the readout architecture.

Several specific and novel tests were developed to experimentally demonstrate the operation and performance of the proposed pixel architecture with non-destructive readout. Since a single pixel of the test structure is instrumented, the PPD and the output stage were operated in different modes to reveal the diverse capabilities of the pixel. Throughout this article, multiple pixel measurements refer to the repetition of measurements on the same single pixel.
\subsection{Multiple non-destructive measurements}
The first step involved characterizing the noise reduction capability with varying numbers of non-destructive samples (NSAMP). To achieve this, the readout stage was isolated from the PPD by maintaining H1 and H2 at a low voltage to mitigate potential noise sources from charge collection. Multiple pixel measurements were recorded in this mode, where no charge contribution was expected. This provided noise information from the output transistors and the readout chain. The procedure was iterated for different NSAMP values per pixel. The maximum value of NSAMP was set to get the desired noise performance for peak resolution.

\begin{figure}[h!]
    \centering
    \includegraphics[page=1,width=\linewidth]{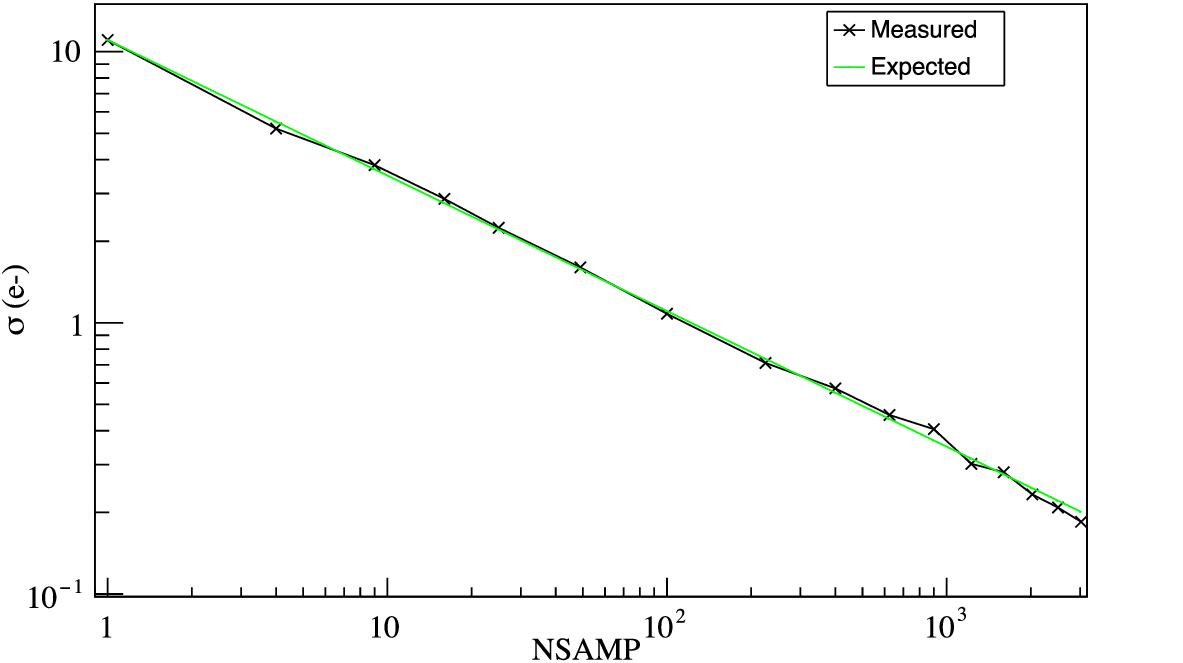}
    \caption{Noise as a function of NSAMP. The black line with asterisks is the measured noise, while the green line is the theoretical expected value.}
    \label{fig:NSAMP_noise_scan}
 \end{figure}

Fig.~\ref{fig:NSAMP_noise_scan} displays the noise standard deviation ($\sigma$) for the group of pixels at each NSAMP value. The noise, presented in equivalent carrier units using an absolute calibration method (described in the following section), is depicted by the black line with asterisks. The green line represents the theoretically expected noise starting from the first measurement at NSAMP $= 1$. 
As per the $1/\sqrt{N}$ law, where N is the number of measurements, the noise reduction should follow a decreasing trend with increasing NSAMP. 
As evident in Fig.~\ref{fig:NSAMP_noise_scan}, the measured noise aligns with this behavior, up to the maximum tested NSAMP of 3025. The last point indicates a deep sub-electron noise performance of 0.18$e^-$, crucial for resolving single electrons in subsequent tests. The integration time employed in this case is 4.6 $\mu$s. The single sample noise, approximately $11e^-$, is higher than expected, which can be attributed to the lack of optimization of the testing setup. Preliminary measurements on standalone transistors within the chip demonstrated lower noise performance.

Using the technique aforementioned, a square signal is applied to V$_{\textbf{ref}}$. This signal amplified is measured at the input of the ADC, allowing to compute the gain in terms of ADUs/V, G$_v=26.6$ADUs/$\mu$ V. Finally, the noise is measured in the flat areas of this square signal, above and below, to obtain the noise of the stand-alone transistor. The noise of the standalone MA source follower transistor was measured at $0.2$e$^-$ using this technique.
\subsection{Charge transfer process in the output stage}
After demonstrating that the noise can be reduced with the number of samples NSAMP, another key parameter of the output stage is the transfer efficiency between the SG and the SN nodes. Once the noise is reduced below the expected signal from a single carrier, any charge collected by the readout stage should be seen as a discrete jump in a temporal series of consecutive measurements\cite{quanta}. If the charge is not dumped, the consecutive pixels should give the same discretized charge information. This method was previously explored in \cite{sensei-single-e-events} to characterize the output stage of the Skipper-CCD and CMOS sensors \cite{quanta}. Following this concept, multiple measurements are taken by repeating the voltage sequence to move the charge back and forth from the SN to SG without dumping the charge into the drain contact. To prevent any extra noise contribution the output stage is isolated from the PPD keeping H1 and H2 at low potential. The metal layers covering the readout output stage prevent external light contributions during this test.

\begin{figure}[h!]
    \centering
    \includegraphics[page=1,width=\linewidth, trim={0 0 0 0}, clip]{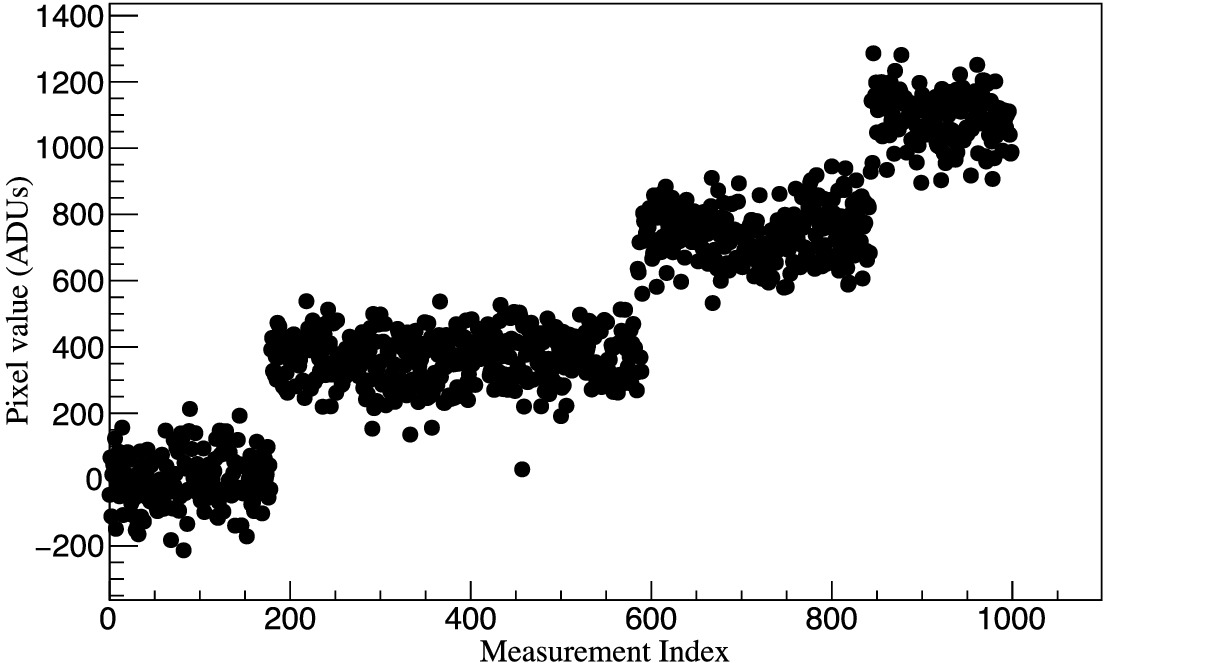}
    \caption{Continuous measurements charge packet in the output stage when isolated from the PPD. Each measurement corresponds to an average of NSAMP $= 3025$ samples. Single electron steps are observed.}
    \label{fig:signal_hist_e}
\end{figure}

Fig.~\ref{fig:signal_hist_e} shows an example of the collected measurements grouped in pixels (each point in the figure) with NSAMP $= 3025$ samples to obtain deep sub-electron noise. The pixel values are in Analog to Digital Units (ADUs). The jumps observed in the series have similar amplitude suggesting quantized increments from the successive charge collected. The points remain constant until the next electron is collected. 
A positive dependence of the rate of collected carriers in the output stage with the biasing current of the source follower transistor was observed, suggesting transistor luminescence \cite{janesick2001scientific}. The biasing current used here is larger than the target design parameter, expecting a further reduction of this rate in future tests.

The same Fig.~\ref{fig:signal_hist_e} is used to compute quantitative preliminary results of the output stage. The sense node sensitivity, G$_e$, can be calibrated with the discrete jumps. The 1e$^-$ jumps correspond to approximately G$_e=350 ADUs$ so the sensitivity of the node is $S=G_e/G_v=13.15\mu$V/e$^-$.
At the same time, the dark current (DC) rate generated at the output stage can be estimated as the 3e$^-$ total jumps in the Fig.~\ref{fig:signal_hist_e} divided by the total measured time, resulting DC rate is below 0.02e$^-$/s.
\subsection{PPD charge collection and transfer to the output stage}
\label{subsec:ppd_charge_collection}
In the final stage of characterization, the sensor's response to an external signal is determined. During this process, the chip is exposed to light for a specific duration while the output stage is maintained at a steady state. Synchronized with the pixel's control sequence, the LED is turned off, while the charge is transferred from the PPD to the output stage for readout. The initial readout corresponds to the light collected by the pixel. To calibrate the pixel reference value for measurements with no charge, multiple readings of the empty PPD were conducted before light exposure. The average of these readings is subtracted from the pixel value with light.

\begin{figure}[h!]
    \includegraphics[page=1,width=\linewidth, trim={0 0 0 0}, clip]{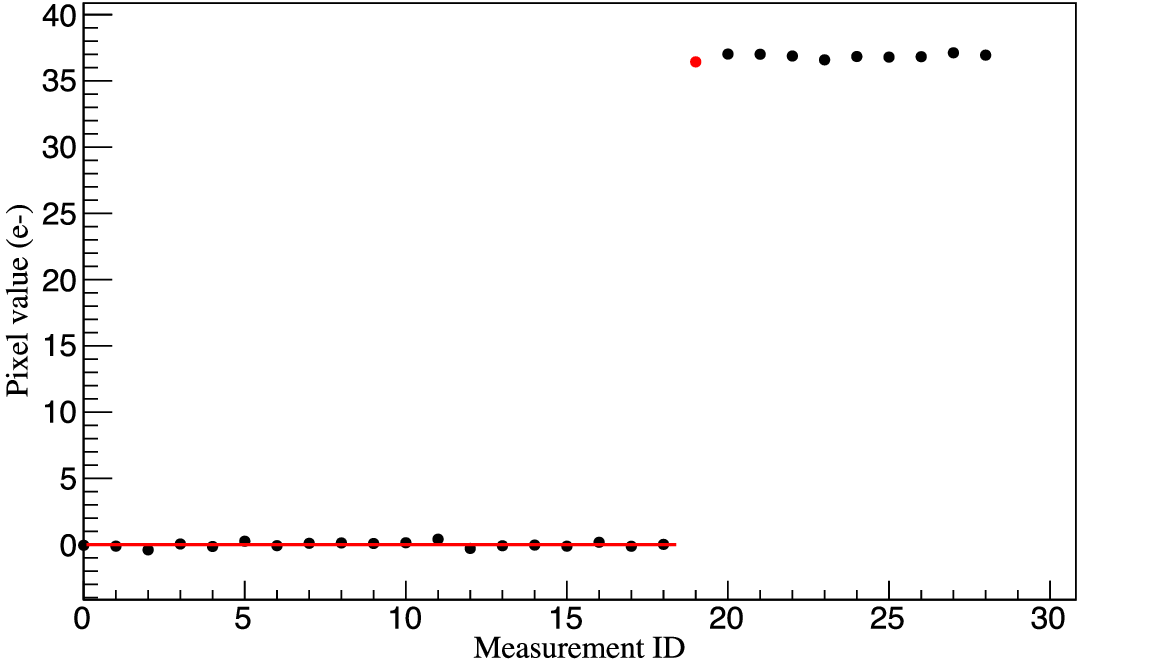}
    \centering
    \caption{The red dot represents the pixel value when the structure is exposed to light, and the line is the average value of the pixels used as the reference value for empty pixels. After ID=18, charge is maintained (no charge dumping).}
    \label{fig:LED_step}
\end{figure}

Fig.~\ref{fig:LED_step} shows one set of measurements following this protocol. The initial nineteen samples are acquired before light exposure, and their average value serves as the reference for empty pixels. Subsequently, the LED is activated for a designated exposure time, and multiple measurements of the collected charges are recorded before charge dumping. The first measurement after LED exposure is utilized as the light-collected information, adjusted by subtracting the reference value from zero-charge pixels. This process is iterated multiple times to gather statistics from the single pixel.

\begin{figure}[h!]
    \centering
    \includegraphics[width=\linewidth, trim={0 0 0 0}, clip]{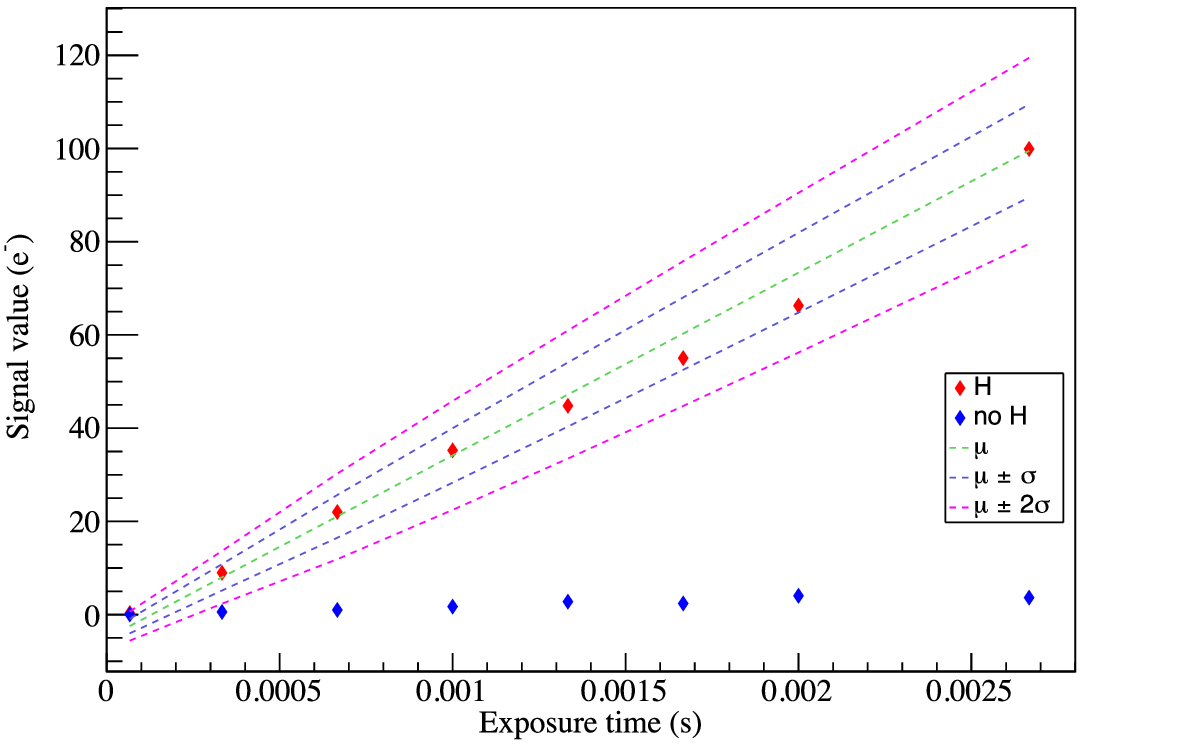}
    \caption{Light response of the pixel: charge transfer to the output stage (red dots) and H clocks as barriers isolating the PPD (blue dots). The dashed lines represent the fitted mean ($\mu$) and $\sigma$ bounds from the Poisson statistics.}
    \label{fig:response_to_light}
\end{figure}

To plot Fig.~\ref{fig:response_to_light}, the above experiment is repeated with varying exposure times ranging up to $2.66$ms. The red dots in the plot show the measured signal, while the blue data were taken with the H clocks low to block the charge flow from the PPD to the output stage. Each data point represents the average value of the collected charge from ten exposures, utilizing non-destructive measurement with NSAMP = 3000. 

These results illustrate that, under normal operation, there is successful charge collection in the PPD, which is efficiently transferred through the gates to the output stage. A Poisson distribution is expected for each exposure time, consistent with photon arrival statistics, with $\sigma=\sqrt{\mu}$. The data points fall within a 1$\sigma$ bound indicated with dashed lines. When the clocks act as a barrier, the PPD remains isolated from the output stage. However, a slight dependence on exposure time (approximately 3e$^-$ at 2.66 ms) is observed for the latter scenario, attributed to light directly collected by the output stage. These preliminary tests utilized a standard, non-calibrated LED light source.

Also, the dark current of the PD was measured for similar exposure times resulting in one-sigma upper bound of $54$e$^-$/s compatible with almost no charge collected for the tests presented here.
\subsection{Single photon counting}
\begin{figure}[t!]
\centering
\includegraphics[width=\linewidth]{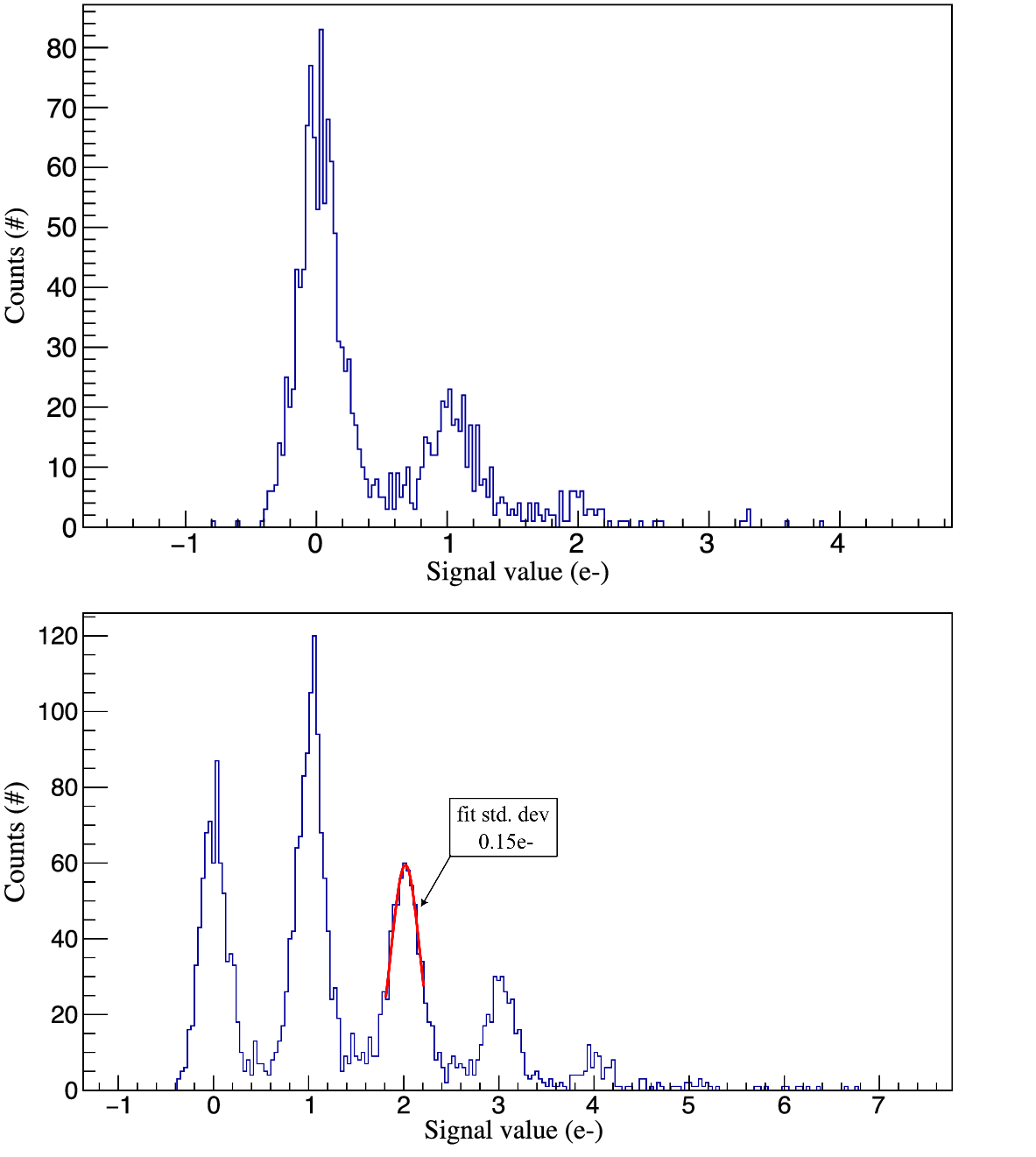}
\caption{Charge discretization for two different LED time exposures, $33\mu$s (top) and $100\mu$s (bottom). The Poisson goodness-of-fit values are $1.002$ and $1.034$ respectively. The bottom figure also shows a normal distribution fit to the $2e^-$ peak, resulting in a noise deviation of $0.15e^-$.}
\label{fig:hist_in_electrons}
\end{figure}

The experiment performed in section \ref{subsec:ppd_charge_collection} is repeated 3000 times to evaluate the pixel architecture's single electron counting capability when exposed to light. Fig.~\ref{fig:hist_in_electrons} shows a histogram of the measured charge packets for (top) $33 \mu$s and (bottom) $100 \mu$s of LED exposure. Both histograms clearly show the single photon counting resolution with very deep subelectron noise. 
Intermediate events between peaks could be attributed to charge generation in the output stage during the readout process of the pixel, producing intermediate charge levels. Some preliminary measurements suggest this charge generation is correlated with the biasing current of the source follower transistor and will be part of future analysis and optimization of the sensor.

The ratio between the mean and variance of these distributions are $1.002$ and $1.034$, respectively, evidencing good matching between the expected Poisson statistical behavior of the photons arriving to the PPD. The bottom of Fig.~\ref{fig:hist_in_electrons} also shows a normal distribution fitted to the $2e^-$ peak, the fitted noise is $0.15e^-$, matching the expected noise value ($0.18e^-$) for a 3000 NSAMP value, from previous measurements, as indicated in Fig.~\ref{fig:NSAMP_noise_scan}. A standard calibration procedure was applied, other methods \cite{Starkey-PCH,hendrickson-PCH-EM} could also be used. This was repeated for all the peaks in the image, presenting a similar standard deviation.
\section{Discussion and future work}
To the best of the authors' knowledge, the results presented in this paper mark the first experimental demonstration of a CMOS image pixel with non-destructive readout achieving deep sub-electron noise, enabling single electron and photon detection. Various techniques and measurements were developed, yielding promising results for the continued advancement of this technology. Table \ref{Table_main_components} summarizes the most important pixel design parameters, readout preliminary results, and principal gate operating voltages.

Future endeavors will focus on optimizing the pixel's performance, fully instrumenting the entire matrix, and measuring standard CMOS performance parameters, such as lag, full well, PD dark current, etc. The pixel utilized in this study represents only one of six different fabrication splits, each with distinct doping profiles, obtained from the initial prototype run. Additionally, the $200\times 200$ pixel matrix is divided into groups of $10$ columns, with each of these $20$ groups featuring different pixel characteristics (e.g., transfer gate width, length, etc.) or flavors.

The assessment of different splits and flavors, coupled with the spatial information provided by the two-dimensional pixel array, will aid in optimizing several parameters, including: 1) Single sample noise improvement: Preliminary measurements of the pixel output transistor noise suggest that the total noise before averaging could be significantly reduced.
2) Detailed study of transfer inefficiencies observed in the output stage of the pixel.
3) Utilization of alternative designs for the dump gate and the drain bias voltage structures to optimize the charge dumping process in the output stage.
4) Reduction of significant intrinsic charge generation observed at the output stage by lowering the biasing current of the output transistor. Applying this reduced current biasing to the pixel matrix is expected to substantially decrease charge generation while maintaining similar performance.
5) Exploration of operation at different temperatures to assess sensor performance at higher and room temperatures.
6) Calibration of different amplifiers, exploration of performance, assessment of correlated noise contributions and crosstalk, and implementation of mitigation techniques. 
\begin{table}[h!]
\centering
\begin{tabular}{|l|l|lll}
\cline{1-2}
Fabrication process        &  TS18 180nm   &  &  &  \\ \cline{1-2}
Pixel size/pitch        &  15$\mu$m   &  &  &  \\ \cline{1-2}
FSI architecture        & yes, current &  &  &  \\ \cline{1-2}
BSI architecture        & yes, back treatment needed &  &  &  \\ \cline{1-2}
Fill factor             &  46\% FSI   &  &  &  \\ \cline{1-2}
Gain  $[\mu V/e^{-}]$   &  13.15   &  &  &  \\ \cline{1-2}
DSI integ. time ($t_i$) [$\mu$s] (not optimized) &  4.66   &  &  &  \\ \cline{1-2}
Readout time per pixel &  $\propto 2t_i$NSAMP   &  &  &  \\ \cline{1-2}
Minimum noise achieved [e$^-$] &   0.15 (with NSAMP = 3025) &  &  &  \\ \cline{1-2}
Voltages H1 [V]&  [$0; 4$]   &  &  &  \\ \cline{1-2}
Voltages H2 [V]&  [$0; 4$]   &  &  &  \\ \cline{1-2}
Voltages SG [V]&  [$0; 2.3$]   &  &  &  \\ \cline{1-2}
Voltages OG [V]&  [$0.7; 1.9$]  &  &  &  \\ \cline{1-2}
Sense node coupling &  floating gate  &  &  &  \\ \cline{1-2}
\end{tabular}
\caption{Main design parameters and preliminary results summary.}
\label{Table_main_components}
\end{table}
\section{Conclusions}
The initial results of a pinned photodiode (PPD) fabricated in CMOS technology with a Skipper output stage showcased several key findings.
It was demonstrated that charge collection in the PPD was efficiently transferred to the output stage and successfully read out.
The PPD was effectively isolated from the output stage via the transfer gates.
The multiple non-destructive readouts of the Skipper output stage resulted in the expected $1/\sqrt{N}$ noise reduction, making it possible to achieve a sub-electronic noise level of $0.15e^-$.
Single electron counting capability was also demonstrated by the observation of discrete steps at the output stage.
Finally, statistical data collection with charged pixels was found to follow the expected Poisson distribution.

These results demonstrate that the development of a Skipper-in-CMOS image sensor in a commercial CMOS process has great potential for applications that require two-dimensional imaging with single electron counting capability using a highly parallelized architecture to reduce the readout time.
\section{Acknowledgments}
This manuscript has been authored by Fermi Research Alliance, LLC under Contract No. DE-AC02-07CH11359 with the U.S. Department of Energy, Office of Science, Office of High Energy Physics. This research has been partially supported by Guillermo Fernandez Moroni's DOE Early Career research program.

\section{Author Declarations and Contributions}
The authors have no conflicts to disclose. Conceptualization: FF, JE, MSH, AD, CK, AF.
Supervision, Project Administration: DB, GFM, SL, FF, BP, LR.
Funding acquisition JE, FF, AD, DB, GFM.
Investigation, Methodology, Formal Analysis, Visualization (Pixel Simulations): AB, MSH, JS.
Investigation, Methodology, Formal Analysis (ASIC design): BP, FAB, LR, AG, MSH.
Investigation, Methodology, Formal Analysis, Data Curation, Visualization (Instrumental development and sensor characterization): AJL, GFM, MSH, FAB, CCB, FC.
Writing – original draft: AJL.
Writing – review \& editing: DB, FF, GFM, FC.
\bibliographystyle{IEEEtran}
\bibliography{IEEEabrv, main.bib}

\end{document}